\documentstyle[twoside,fleqn,espcrc2]{article}

\def\eq#1{Eq.~(\ref{#1})}
\newcommand{\secn}[1]{Section~\ref{#1}}
\newcommand{\bra}[1]{\langle{#1}|}
\newcommand{\ket}[1]{|{#1}\rangle}

\newcommand{\nl}{\nonumber \\}

\def\beq{\begin{equation}}
\def\eeq{\end{equation}}
\def\beqa{\begin{eqnarray}}
\def\eeqa{\end{eqnarray}}

\newcommand{\sect}[1]{\setcounter{equation}{0}\section{#1}}

\newcommand{\as}{\alpha_s}
\renewcommand{\b}{\beta}

\newcommand{\m}{\mu}
\newcommand{\g}{\gamma}
\newcommand{\e}{\epsilon}

\newcommand{\AmS}{{\protect\the\textfont2
  A\kern-.1667em\lower.5ex\hbox{M}\kern-.125emS}}

\title{The resummed quark form factor in dimensional regularization}

\author{L. Magnea\address{Dipartimento di Fisica Teorica,
Universit\`a di Torino \\
and I.N.F.N., Sezione di Torino\\
Via P.Giuria 1, I--10125 Torino, Italy}}

\begin{document}

\begin{abstract}

The resummed expression for the quark form factor illustrates the fact
that dimensional continuation provides a regularization not only for
ultraviolet and infrared singularities of fixed order QCD amplitudes,
but also for the Landau pole arising in resummed calculations.
Explicit renormalization group invariant analytic expressions for the
logarithm of the form factor are given up to two--loop order in the
QCD $\b$ function.

\end{abstract}

\maketitle

\section{Introduction}
\label{intr}

The resummation of perturbation theory for the form factor of charged
gauge particles has a long history, and has gone through several
levels of refinement, starting with the seminal work of
Sudakov~\cite{suda}.  The complete exponentiation of the form factor
can be achieved by deriving and solving an evolution equation
governing its energy dependence, as was done in Ref.~\cite{evol}, by
several authors, using different techniques. An instructive
introduction to some of these techniques can be found in
Ref.~\cite{collrev}. The usefulness of dimensional regularization for
the implementation and the solution of the evolution equation was
first noticed in Ref.~\cite{us}, where an explicit exponentiated
solution for the form factor, directly comparable with diagrammatic
calculations, was given, and also the ratio of the timelike to the
spacelike form factor was computed.  This solution expresses the
logarithm of the form factor in terms of integrals over the scale
$\m^2$ of the running coupling, with an integration region extending
all the way to $\m^2 = 0$, as always in resummed expressions for QCD
amplitudes and cross sections. In $d = 4$, these integrals are
ill--defined because of the Landau pole singularity in the running
coupling, and typically the ensuing ambiguity in the resummed
expression is taken as an estimate for the size of nonperturbative,
power--suppressed effects~\cite{bene}. More recently \cite{me}, I have
shown that dimensional continuation provides a natural and gauge
invariant way to regularize the Landau singularity; the integrals over
the scale of the coupling can then be explicitly evaluated in terms of
analytic functions of the coupling and of the space--time dimension,
which are renormalization group invariant by inspection to the desired
accuracy in the QCD $\b$ function.  Here, I will briefly review the
results of Ref.~\cite{me}, emphasizing the special role played by
dimensional regularization and the fate of the Landau pole.

\section{Resumming the form factor with dimensional regularization}
\label{hist}

For definiteness, consider the timelike electromagnetic quark form
factor in dimensionally regularized massless QCD, defined by
\beqa 
& & \hspace{-9mm} \Gamma_\mu (p_1, p_2; \mu^2, \e) = \bra{0}
J_\mu (0) \ket{{p_1,p_2}} \label{def} \\
& = & - {\rm i} e e_q ~\overline{v}(p_2) \gamma_\mu
u(p_1) ~\Gamma \left( \frac{Q^2}{\mu^2}, \as(\mu^2), \e \right)~, \nonumber
\eeqa
where $J_\mu$ is the electromagnetic current, $Q^2 = (p_1 + p_2)^2$,
and $\e = 2 - d/2 < 0$ to regulate IR divergences in the renormalized
theory.  As described in Ref.~\cite{collrev}, the $Q^2$ dependence of
the form factor is determined by an evolution equation of the form
\beqa
& & \hspace{-9mm} Q^2 \frac{\partial}{\partial Q^2} \log \left[\Gamma \left( 
\frac{Q^2}{\mu^2}, \as(\mu^2), \e \right) \right] \label{eveq} \\
& = & \frac{1}{2} \left[ K \left(\e, \as(\mu^2) \right) + 
G \left(\frac{Q^2}{\mu^2}, \as(\mu^2), \e \right) \right]~. \nonumber
\eeqa
The function $G$ contains the $Q^2$ dependence and is finite in the
limit $\e \to 0$, while the function $K$ is a pure counterterm;
furthermore, $K$ and $G$ must renormalize additively with the same
anomalous dimension $\g_K$ to preserve the renormalization group
invariance of the full form factor: thus, $d G/d \ln \m = - d K/d \ln
\m = \g_K(\as)$. The functions $K$, $G$ and $\g_K$ are perturbatively
calculable and known to two loops.

Dimensional regularization affects \eq{eveq} in two crucial
ways. First, renormalization group evolution is dictated by the
$d$--dimensional $\b$ function,
\beq
\b(\e,\as) = \mu \frac{\partial \as}{\partial \mu} = - 2 \e \as +
\hat{\b} (\as)~, 
\label{beta}
\eeq
where $\hat{\b} (\as)$ is the usual $\b$ function in $d = 4$. As a
consequence, for example, the one--loop running coupling takes the
form
\beqa
& &  \hspace{-6mm} \overline{\alpha}\left(\frac{\mu^2}{\mu_0^2},
\as(\mu_0^2),\e\right) ~=~
\as(\mu_0^2) \left[\left(\frac{\mu^2}{\mu_0^2}\right)^\e \right. 
\label{loalpha} \\ & & - \left. \frac{1}{\e}
\left(1 - \left(\frac{\mu^2}{\mu_0^2}\right)^\e \right) \frac{b_0}{4 \pi}
\as(\mu_0^2) \right]^{-1}~, \nonumber
\eeqa
with $b_0 = (11 C_A - 2 n_f)/3$.  The second effect of dimensional
regularization is that one may explicitly solve the evolution equation
(\ref{eveq}) with the simple boundary condition
\beq
\Gamma \left(0,\as(\mu^2),\e \right) = 1~.
\label{incon}
\eeq
This can be seen by considering the perturbative expansion for
$\Gamma(Q^2)$, in which each term must be proportional to a positive
integer power of $(\m^2/(-Q^2))^\e$, or alternatively by noting that
the $d$--dimensional running coupling in \eq{loalpha} {\it vanishes},
when $\m^2 \to 0$, as $\m^{- 2 \e}$. One can then solve \eq{eveq}
obtaining
\beqa
& & \hspace{-9mm} \Gamma \left( \frac{Q^2}{\mu^2}, \as(\mu^2), \e \right) =
\exp \left\{ \frac{1}{2} \int_0^{- Q^2} \frac{d \xi^2}{\xi^2} \right. 
\label{sol} \\ 
& & \hspace{-9mm} \left[ K \left(\e, \as(\mu^2) \right) + G \left(-1, 
\overline{\alpha} \left(\frac{\xi^2}{\mu^2},\as(\mu^2),\e \right), 
\e \right) \right. \nl
& + & \left. \left. \frac{1}{2} \int_{\xi^2}^{\mu^2} 
\frac{d \lambda^2}{\lambda^2} \gamma_K \left(\overline{\alpha} 
\left(\frac{\lambda^2}{\mu^2},\as(\mu^2), \e \right) \right) \right] 
\right\}~. \nonumber
\eeqa
\eq{sol} was used in Ref.~\cite{us} to derive an explicit
representation for the ratio $\Gamma(Q^2)/\Gamma(-Q^2)$, which is
phenomenologically relevant beacuse it enters directly the resummed
partonic cross section of the Drell--Yan process \cite{dysum}.

A further illustration of the power of dimensional regularization is
given by the computation of the ``counterterm'' function $K(\e, \as)$
\cite{me}.  In a minimal scheme, $K$ depends on $\m$ only through the
coupling $\as(\m^2)$; thus, its RG equation becomes a recursion
relation, expressing all higher order poles in terms of the
perturbative coefficients of the simple pole, these in turn being
determined by the anomalous dimension $\g_K$.  In particular, writing
\beqa
K(\e,\as) & = & \sum_{m = 0}^\infty {\cal K}_m (\e, \as)~, \label{serpol} \\
{\cal K}_m (\e, \as) & = & \sum_{n = 1}^\infty K_n^{(n + m)} 
\left(\frac{\as}{\pi}\right)^{n + m} \frac{1}{\e^n}~, \nonumber
\eeqa
one observes that all leading poles ($m = 0$) are determined by one
loop calculations, while in general the coefficients of the poles
contributing to ${\cal K}_m$ require a calculation to $m + 1$
loops. It turns out \cite{me} that the recursion relation determining
the coefficients $K_n^{(m)}$, which depends on the $\b$ function given
by \eq{beta}, can be solved completely, including {\it all} orders in
$\hat{\b}(\as)$. Furthermore, all the resulting series of poles ${\cal K}_m
(\e, \as)$ can be summed, and yield analytic functions of $\as$ and
$\e$ that are {\it regular} as $\e \to 0$ for $m > 0$. The only
singularity in the resummed expression for $K(\e, \as)$ is logarithmic
and completely determined by the one--loop $\b$ function and the
one--loop coefficient in the anomalous dimension $\g_K$.
Specifically, including only one--loop results, one finds \cite{us,me}
\beq
K_n^{(m)} = \frac{1}{2 m} \left(- \frac{b_0}{4} \right)^{n - 1} 
\gamma_K^{(m - n  + 1)}~,
\label{ols}
\eeq
which is exact for $n = m$ and implies
\beq
{\cal K}_0 (\e, \as) = \frac{2 \gamma_K^{(1)}}{b_0} 
\ln \left(1 + \frac{b_0 \as}{4 \pi \e}
\right)~.
\label{klp}
\eeq
Notice that the function ${\cal K}_0 (\e, \as)$ has a cut in the $\e$
complex plane running from $\e = - b_0 \as/(4 \pi)$ to $\e = 0^-$.
This cut is a direct consequence of the Landau singularity in the
one--loop running coupling, as I will discuss in the next section.

\section{The fate of the Landau pole}
\label{land}

To understand why dimensional continuation allows for such explicit
evaluations of resummed quantities in QCD, one may consider the
one--loop $\b$ function at $\e < 0$, as given by \eq{beta}. It is
apparent that instead of the usual (double) zero at $\as = 0$, the
dimensionally continued $\b$ function has two distinct zeroes, one at
the origin and one located at $\as = - 4 \pi \e/b_0$. For $\e < 0$,
this second zero is the asymptotically free one, while at the origin
in the $\as$ plane the $\beta$ function vanishes with positive
derivative, as it would in a QED-like theory. This is a different way
of expressing the fact that the running coupling vanishes like a power
of the scale for $\e < 0$, as pointed out in \secn{hist}\footnote{It
should be noted that the use of dimensional regularization in this
spirit is a customary tool in statistical field
theory~\cite{paretal}}. The presence of the second, asymptotically
free zero of the $\b$ function is manifest in the explicit expression,
\eq{loalpha}, for the one--loop running coupling.  In fact,
\eq{loalpha}, just like the four--dimensional running coupling, has a
simple pole in the $\m^2$ complex plane, located at
\beq
\mu^2 = \Lambda^2 \equiv Q^2 \left(1 + \frac{4 \pi \e}{b_0 \as(Q^2)}
\right)^{-1/\e}~,
\label{lapo}
\eeq
which can be used to define the coupling, just as in conventional
dimensional transmutation,
\beq
\as(Q^2) = \frac{4 \pi \e}{b_0 \left[ \left( \frac{Q^2}{\Lambda^2} 
\right)^\e - 1 \right]}~.
\label{asq}
\eeq
What has changed as an effect of dimensional continuation is the fact
that now the Landau pole is not necessarily located at real values of
the renormalization scale. In fact, for $\e < - b_0 \as/(4 \pi)$, $\e
\neq - 1/n$, $\Lambda^2$ acquires a nonvanishing imaginary part. For
such (large) values of the space--time dimension, the coupling
decreases smoothly to zero, starting from the boundary value
$\as(Q^2)$, without encountering any singularity for real values of
the scale. On the other hand, when the dimension of space time is
sufficiently close to $d = 4$, the Landau pole migrates to the real
axis in the $\m^2$ complex plane, which is also the integration
contour for resummed expressions such as \eq{sol}. As a result, the
analytic expressions obtained by evaluating the integrals will develop
a cut, which might appropriately be called Landau cut.

One sees that dimensional continuation succeds in regularizing the
Landau singularity, arising from resummation, much in the same way as
it regularizes ultraviolet and infrared divergences in fixed order
perturbative calculations: instead of an ill--defined expression, we
will now find an analytic function of the coupling and of the
space--time dimension, with a specified singularity (in this case a
cut) when the physical value $d = 4$ is approached. Of course, the
singularity has not been eliminated, since it has a physical meaning
and is ultimately related to confinement; rather, the singularity is
now parametrized in a gauge--invariant manner, which might provide
insights into nonperturbative corrections to resummed physical
quantities, if the method can be applied to infrared safe observables.

\section{One--loop analytic resummation}
\label{onel}

Given the discussion in the previous section, one expects that it
should be possible to evaluate explicitly the integrals in \eq{sol}.
Furthermore, using for the running coupling the solution of the RG
equation to a given order in $\hat{\b}(\as)$, one expects the result to be
RG invariant, {\it i.e.} independent of the renormalization scale
$\m^2$, to the same accuracy. One can readily verify these claims at
the one--loop level. Using \eq{loalpha}, and changing variables
according to
\beq 
\lambda^2 \to z = \left(\frac{\m^2}{\lambda^2}\right)^\e - 1~, 
\label{chvar}
\eeq
the anomalous dimension integral is easily performed, yielding a
logarithm.  The logarithm is not integrable over the scale $\xi^2$
because of a singularity at $\xi^2 = 0$, which however is cancelled by
the $\xi$--independent contribution of the counterterm function $K$,
\eq{klp}.  Inserting the values of the one--loop coefficients of $K$,
$\g_K$ and $G$, and defining for simplicity
\beq
a(\m^2) = \frac{b_0}{4 \pi} \as(\m^2)~, 
\label{am}
\eeq
one finds
\beqa
& & \hspace{-7mm} \log \Gamma \left(\frac{- Q^2}{\mu^2}, \as(\mu^2), 
\e \right) = - \frac{2 C_F}{b_0} \label{ln1} \\
& & \times \left\{ \frac{1}{\e} ~{\rm Li}_2 \left[
\left(\frac{\mu^2}{Q^2} \right)^\e \frac{a(\mu^2)}{a(\mu^2) + \e} \right]
\right. \nl
& & \left. - ~C(\e) ~\ln 
\left[1 - \left(\frac{\mu^2}{Q^2} \right)^\e 
\frac{a(\mu^2)}{a(\mu^2) + \e} \right] \right\}~,
\nonumber
\eeqa
where for clarity I considered the spacelike rather than the timelike
form factor, and
\beq
C(\e) = (3 - \e (\zeta(2) - 8))/2 + {\cal O} (\e^2)
\label{ce}
\eeq
arises from the one--loop contribution to the function $G$.

RG invariance of \eq{ln1} is readily verified by reexpressing the
running coupling in terms of $\as(Q^2)$, using \eq{loalpha}. One
finds that the $\m^2$ dependence cancels, obtaining
\beqa
& & \hspace{- 7mm} \log \Gamma \left(\frac{- Q^2}{\mu^2}, \as(\mu^2), 
\e \right) =~ \log \Gamma \left(- 1, \as(Q^2), \e \right) \nl
& & =  - \frac{2 C_F}{b_0} ~\left\{ \frac{1}{\e} ~{\rm Li}_2 \left[
\frac{a(Q^2)}{a(Q^2) + \e} \right] \right. \label{rgi1} \\
& & + \left. C(\e) ~\log \left[1 + \frac{a(Q^2)}{\e}\right] \right\}~.
\nonumber
\eeqa
\eq{rgi1} is a striking illustration of the power of dimensional
continuation; the resummed form factor has a simple analytic
structure, characterized by a cut (the ``Landau cut''), which can be
taken to run from $\e = - a(Q^2)$ to $\e = 0^-$, as expected; one can
reexpand \eq{rgi1} in powers of $\as(Q^2)$ for fixed $\e$ to recover
known perturbative results, or to generate the coefficients of all
leading and next--to--leading infrared and collinear poles of the form
factor; on the other hand, one can examine the behavior of the
resummed expression in the vicinity of the physical point $\e = 0$;
one finds then
\beqa
& & \hspace{-7mm} \log \Gamma \left(- 1, \as(Q^2), \e \right) =
\frac{2 C_F}{b_0} \Bigg[ - \frac{\zeta(2)}{\e} + \frac{1}{a(Q^2)} 
\Bigg. \nl & & \hspace{-7mm} + \Bigg. 
\left( \frac{1}{a(Q^2)} - \frac{3}{2} \right) \log \left(\frac{a(Q^2)}{\e}
\right) + {\cal O} (\e) \Bigg]~.
\label{eto01}
\eeqa
One observes that the resummation of all leading poles in the
logarithm of the form factor gives just a single pole, with a residue
independent of the coupling and of the energy, up to logarithmic
corrections. As will be seen below, one may conjecture that this is in
fact the only pole in the complete perturbative resummation of the
logarithm of the form factor.  It is also worth noticing that, in the
vicinity of $\e = 0$, the form factor contains a {\it finite} term of
the form $\exp (c/\as)$, which would correspond to a power--behaved
contribution of the type $(\Lambda^2/Q^2)^{-c}$ in the
four--dimensional theory. Although in the present case this term is of
no direct physical significance, being associated with an IR divergent
quantity, it is encouraging to see a term of this kind arising in a
gauge--invariant way from the present formalism. Finally, it should be
mentioned that \eq{rgi1} can be tested by computing the ratio of the
timelike to the spacelike form factor, which is found to agree with
the results of \cite{us}.

\section{Two--loop analytic resummation}
\label{twol}

To generalize to higher perturbative orders the calculation performed
in \secn{onel}, one can take advantage of the fact that integrations
over the renormalization scale can be replaced by integrations over
the coupling itself, using
\beq
\frac{d \mu}{\mu} = \frac{d \as}{\beta(\e, \as)}~,
\label{chva}
\eeq
and truncating the perturbative $\b$ function to the desired order. It
is easy to reproduce the one--loop result, \eq{rgi1}, with this
method.  At two loops, one must change variables in both scale
integrals according to
\beq
\frac{d \mu^2}{\mu^2} = - \frac{d \as}{\as} 
\frac{1}{\e + \frac{b_0}{4 \pi} \as + \frac{b_1}{4 \pi^2} \as^2}~,
\label{chva2}
\eeq
and of course one must include in the calculation the two--loop
coefficients of the functions $K$, $G$ and $\g_K$. All integrations
can be explicitly performed by partial fractioning, and they yield
logarithms and dilogarithms, expressed in terms of the two nontrivial
zeroes of the two--loop $\b$ function.  Writing
\beq
a_\pm = - \frac{b_0}{2 b_1} \left(1 \pm \sqrt{1 - \frac{16 \e b_1}{b_0^2}}
\right)~.
\label{zeroes}
\eeq
one finds
\beqa
& & \hspace{-7mm} \log \Gamma \left(- 1, \as(Q^2), \e \right) =
- \frac{2}{b_1 (a_+ - a_-)} \nl
& & \hspace{-7mm} \times \left\{ \left( G^{(1)}(\e) + 
a_+ G^{(2)}(\e) \right) \log \left(1 - \frac{\as (Q^2)}{\pi a_+} \right)
\right. \nl
& & \hspace{-7mm} + ~2 \left(\gamma_K^{(1)} + a_+ 
\gamma_K^{(2)} \right) \left[
- \frac{1}{4 \e} {\rm Li}_2 \left( \frac{\as(Q^2)}{\pi a_+} 
\right) \right. \nl
& & \hspace{-7mm} + \frac{1}{2 b_1 a_+ (a_+ - a_-)} 
\log^2 \left(1 - \frac{\as(Q^2)}{\pi a_+} \right) \nl
& & \hspace{-7mm} - ~\frac{1}{b_1 a_- (a_+ - a_-)} 
\left( {\rm Li}_2 \left( 
\frac{\pi a_+ - \as(Q^2)}{\pi (a_+ - a_-)} \right) \right. \nl
& & \hspace {-7mm} + ~\log 
\left(1 - \frac{\as(Q^2)}{\pi a_+} \right) 
\log \left(\frac{\as(Q^2) - \pi a_-}{\pi  (a_+ - a_-)} \right) \nl
& & \hspace {-8mm} \left. \left. \left. - ~{\rm Li}_2 
\left( \frac{a_+}{a_+ - a_-} \right)
\right) \right] \right\} ~+ ~\left( a_+ \leftrightarrow a_- \right)~.
\label{rgi2}
\eeqa
It is fairly straightforward to verify that when the two--loop
coefficients $\{ b_1, \g_K^{(2)}, G^{(2)}(\e) \}$ are taken to zero
\eq{rgi2} reduces to \eq{rgi1} . One can also study the behavior of
\eq{rgi2} in the vicinity of $\e = 0$; one finds that, as announced,
the simple pole in the logarithm of the form factor, given in
\eq{eto01}, is {\it not} affected by the inclusion of two--loop
effects, while the logarithmic singularities are enhanced to $\log^2$
strength. This result, together with the all--loop evaluation of the
counterterm function $K$, strongly suggests that the residue of the
simple pole in \eq{eto01} receives {\it no} higher order perturbative
corrections.

It is apparent that the calculation leading to \eq{rgi2} can be
generalized to any number of loops: with the change of variables in
\eq{chva}, one must deal at most with a double integral of a rational
function, which is in general computable in terms of polylogarithms by
means of partial fractioning. While this generalization would be
purely formal at this stage, since the three--loop perturbative
coefficients of the functions $G$ and $\g_K$ are not known at present,
one may rely on the fact that the present method and results can be
systematically improved upon when higher order perturbative
calculations become available.

\vspace{-2pt}
\sect{Outlook}
\label{outl}

I have shown that, in the case of the resummed quark form factor,
dimensional continuation provides a gauge-- and RG--invariant
regularization of the Landau singularity. A generalization of the
present formalism to more complicated QCD amplitudes and cross sections,
and in particular to the resummation of real soft gluon emission,
would be of great theoretical and phenomenological interest, since it
would in principle lead towards the construction of resummed,
RG--invariant partonic cross section, and might provide useful
insights in the nature of nonperturbative corrections to factorization
theorems.  With this in mind, it is encouraging to note that the form
factor plays a key role also in the resummation of real gluon emission
\cite{prep}, and that ``Sudakov'' resummation techniques are available
also for more complicated QCD processes, involving nonsiglet color
exchanges \cite{ttbar}.


\begin{thebibliography}{99}

\bibitem{suda} V.V. Sudakov, {\it Sov. Phys. JETP} {\bf 3} (1956) p. 65.

\bibitem{evol} A.H. Mueller, {\it Phys. Rev.}  {\bf D 20} (1979)
p. 2037; J.C. Collins, {\it Phys. Rev.}  {\bf D 22} (1980) p. 1478;
A. Sen, {\it Phys. Rev.}  {\bf D 24} (1981) p. 3281; G.P. Korchemsky
and A.V. Radyushkin {\it Nucl. Phys.} {\bf B 283} (1987) p. 342.

\bibitem{collrev} J.C. Collins, in {\it Perturbative Quantum
Chromodynamics}, ed. A.H. Mueller, World Scientific, Singapore, 1989,
p. 573.

\bibitem{us} L. Magnea and G. Sterman, {\it Phys. Rev.} {\bf D 42} (1990)
p. 4222.

\bibitem{bene} M. Beneke, {\it Phys. Rep.} {\bf 317} (1999) p. 1, 
{\tt hep-ph/9807443}.
 
\bibitem{me} L. Magnea, {\tt hep-ph/0006255}.

\bibitem{dysum} G. Sterman, {\it Nucl. Phys.} {\bf B 281} (1987)
p. 310; S. Catani and L. Trentadue, {\it Nucl. Phys.} {\bf B 327}
(1989) p. 323; L. Magnea, {\it Nucl. Phys.} {\bf B 349} (1991) p. 703.

\bibitem{paretal} See, {\it e.g.}, G. Parisi, {\it Statistical 
field theory}, Addison Wesley (1988).

\bibitem{prep} A. Bassetto, M. Ciafaloni and G. Marchesini, 
{\it Phys. Rep.} {\bf 100} (1983) p. 201.

\bibitem{ttbar} N. Kidonakis and G. Sterman, {\it Nucl. Phys.} {\bf B
505} (1997) p. 321, {\tt hep-ph/9705234}; R. Bonciani, S. Catani,
M.L. Mangano and P. Nason, {\it Nucl. Phys.} {\bf B 529} (1998)
p. 424, {\tt hep-ph/9801375}.

\end{thebibliography}
\end{document}